\documentclass[preprint2,times,tighten]{aastex62}
\pdfoutput=1 
\usepackage{amsmath,amstext,amssymb}
\usepackage{wasysym} 						
\usepackage{verbatim}
\usepackage{soul}
\usepackage[T1]{fontenc}
\usepackage{apjfonts} 
\usepackage[figure,figure*]{hypcap}
\usepackage{rotating}

\newcommand{\lya}{Ly$\alpha$}

\shorttitle{Turbulent Absorption Lines}
\shortauthors{Buie et al.}

\begin{document}

\title{Interpreting Observations of Absorption Lines in the Circumgalactic Medium with a Turbulent Medium}

\author{Edward Buie II}
\affiliation{Arizona State University School of Earth and Space Exploration, P.O. Box 871404, Tempe, AZ 
85287, USA}
\author{Michele Fumagalli}
\affiliation{
Centre for Extragalactic Astronomy, Durham University, South Road, Durham, DH1 3LE, UK}
\affiliation{Institute for Computational Cosmology, Durham University, South Road, Durham, DH1 3LE, UK}
\affiliation{Dipartimento di Fisica G. Occhialini, Universit\`a degli Studi di Milano Bicocca, Piazza della Scienza 3, 20126 Milano, Italy 
}
\author{Evan Scannapieco} 
\affiliation{Arizona State University School of Earth and Space Exploration, P.O. Box 871404, Tempe, AZ 
85287, USA}

\begin{abstract}
Single-phase photoionization equilibrium (PIE) models are often used to infer the underlying physical properties of galaxy halos probed in absorption with ions at different ionization potentials. 
To incorporate the effects of turbulence, we use the MAIHEM code to model an isotropic turbulent medium exposed to a redshift zero metagalactic UV background, while tracking the ionizations, recombinations, and species-by-species radiative cooling for a wide range of ions. By comparing observations and simulations over a wide range of turbulent velocities, densities, and metallicity with a Markov chain Monte Carlo technique, we find that MAIHEM models provide an equally good fit to the observed low-ionization species compared to PIE models, while reproducing at the same time high-ionization species such as \ion{Si}{4} and \ion{O}{6}. 
By including multiple phases, MAIHEM models favor a higher metallicity ($Z/Z_\odot \approx 40\%$) for the circumgalactic medium
compared to PIE models. 
Furthermore, all of the solutions require some amount of turbulence ($\sigma_{\rm 3D} \geqslant 26\ {\rm km}\ {\rm s}^{-1}$). Correlations between turbulence, metallicity, column density, and impact parameter are discussed alongside mechanisms that drive turbulence within the halo. 

\end{abstract}


\section{Introduction}

In the prevailing cosmological model, the primary driver of structure formation is completely invisible.  Such dark matter encompasses  $\approx 86\%$ of the material in the universe \citep{planck2018}, and interacts purely gravitionally, gathering into bound structures that merge and accrete over time. Such halos of dark matter form the sites of galaxy formation, as gas falls into them, condenses, and cools into the interstellar medium from which stars are then formed.  

The majority of this baryonic material remains almost as invisible as the dark matter. Galaxies themselves occupy only the centers of  halos, while most of the volume is filled with a diffuse `circumgalactic medium' (CGM), which is observable via absorption-line measurements using background quasars \citep[e.g.,][]{tripp2008,chen2010,prochaska2011,tumlinson2013,savage2014,werk2014}, or through emission for example 
in faint X-ray 
\citep[e.g.,][]{gupta2012,anderson2013,miller2015}
or HI \lya\ emission \citep[e.g.,][]{thom2012,cantalupo2014cosmic}.

The limited information we have on the CGM has been difficult to interpret. The pressures derived from equilibrium modeling of cold absorption line systems \citep[e.g.,][]{werk2014cos} are often order of magnitudes lower than those derived from X-ray observations \citep{tumlinson2017circumgalactic}. Moreover, UV absorption studies frequently reveal gas in very different ionization states at coincident velocities along the same line of sight \citep{tripp2008,tripp2011}, leading to the notion of a 
multiphase CGM.

Most recently, the Cosmic Origins Spectrograph (COS) on the Hubble Space Telescope (HST) was used to conduct absorption line studies on low-redshift galaxies \citep[e.g.,][]{prochaska2011, tumlinson2013}. Some of the findings from this COS-Halos survey include declining low ionization state gas as a function of impact parameter and \ion{O}{6} absorption spanning the entire CGM of $L_{*}$ galaxies. However, a puzzling result is the lack of \ion{N}{5} absorption \citep{werk2016ApJ...833...54W}, which is odd given the ionization potentials for \ion{N}{5} and \ion{O}{6}  differ by only 40 eV, being 98 eV and 138 eV, respectively.

Given its multiphase nature, complex modeling is required to simultaneously capture  all the observed features of the CGM. Typically, single-phase photoionization equilibrium (PIE) models are able to reproduce only the low and intermediate ions, while collisional ionization equilibrium models are needed to produce the higher state ions like \ion{O}{6} \citep{tumlinson2017circumgalactic}. However, to have all three of these ionization states present at the same time, more complex non-equilibrium models
are required. Among those, there are models that focus on the gas dynamics that may release ionizing radiation while cooling \citep{2012wakkerApJ...749..157W,mcquinn2018implications}, models with radiative shocks that leave ionized gas in their wakes \citep{2009gnatApJ...693.1514G,lochhaas2018fast}, or models with conductive interfaces that allow the transfer of heat between cold clouds and an ambient condensing hot medium \citep{sembach2003highly,cottle2018column}.

In addition to the CGM being multiphase, \citet{werk2016ApJ...833...54W} find the higher ionization state gas to have turbulent velocities of 50 - 75 km s$^{-1}$ contributing to the line widths of ions, which do not find a natural explanation in PIE models. To explore the dynamics of such turbulence in the CGM, \citet{buie2018modeling} modeled isotropic turbulence using MAIHEM\footnote{http://maihem.asu.edu/} (Models of Agitated and Illuminated Hindering and Emitting Media), in attempts to explain the puzzling presence of \ion{O}{6} but lack of \ion{N}{5} absorption feature. This work showed that turbulence with $\sigma_{\rm 1D} \approx 60$ km s$^{-1}$ replicates many of the observed features within the CGM, such as clumping of low ionization-state ions and the existence of \ion{O}{6} at moderate ionization parameters, over-predicting however the amount of \ion{N}{5}. 

Building on this work, in this paper we conduct a Markov Chain Monte Carlo (MCMC) analysis to investigate whether the MAIHEM models describe accurately COS-Halos data,  and to test how such turbulent velocities would alter the probability distribution function (PDF) of physical parameters of the CGM (such as metallicity) as compared to those found previously using PIE models. In particular, we focus on the analysis of COS-Halos observations presented in  \citet{prochaska2017cos} (hereafter P17). These data were previously analyzed using PIE models which, as stated, are usually able to reproduce the low and intermediate ions but have trouble in matching \ion{O}{6} \citep{lehner2013bimodal} without invoking very-low densities \citep{stern2016}.

The structure of this work is as follows. In \S 2 we discuss the observational data, and in \S 3 we give a brief description of MAIHEM as well as the MCMC code used to conduct our analysis. In \S 4 we show the main results of our analysis, followed by a discussion in \S 5.

\section{Data}
The data used for this analysis is a compilation of absorption line systems probing the CGM of low-redshift galaxies from the COS-Halos survey. Specifically, we model the observations presented in P17, which were previously analyzed using PIE models. These data consist of several ions for multiple absorbers along quasar sightlines that probe $M_{*} = 10^{9.5} - 10^{11.5}$ \textit{M}$_{\odot}$ galaxies out to an impact parameter $b \approx 150$ kpc \citep{tuml2013ApJ...777...59T}. The reader is encouraged to refer to previous work by the COS-Halos survey for additional detail on the data collection and processing. 

Here, we study the same COS-Halos sightlines that were analyzed in P17, for a total of 44 low-redshift ($z$ $\lesssim$ 0.5) systems. These sightlines exhibit absorption from a mixture of ionization states that include \ion{C}{2}, \ion{C}{3}, \ion{Fe}{2}, \ion{Fe}{3}, \ion{Mg}{2}, \ion{N}{2}, \ion{O}{1}, \ion{O}{6}, \ion{Si}{2}, \ion{Si}{3}, \ion{Si}{4}, and \ion{S}{3}. As we now have the ability to model a multiphase medium, we also include \ion{Si}{4} and \ion{O}{6} the column density of which were not incorporated in the analysis of P17. 

We conduct the MCMC analysis on the entire sample, finding however that only $32/44$ systems converge on solutions with the MAIHEM model. Systems that are unable to find solutions have either all upper or lower limits, or have only accurate column density measurements for a single ion. When drawing comparisons with P17, we thus limit our sample to only include the 32 systems with converged solution.

\section{Methods}
\subsection{MAIHEM Models}
We use the MAIHEM code to model ionic fractions in an isotropic turbulent medium. This cooling and chemistry package is a modified version of the open-source hydrodynamics code FLASH (Version 4.4) \citep{fryxell2000flash}, which explicitly tracks the reaction network of 65 ions, including: hydrogen (\ion{H}{1} and \ion{H}{2}), helium (\ion{He}{1} -- \ion{He}{3}), carbon (\ion{C}{1} -- \ion{C}{6}), nitrogen (\ion{N}{1} -- \ion{N}{7}), oxygen (\ion{O}{1} -- \ion{O}{8}), neon (\ion{Ne}{1} -- \ion{Ne}{10}), sodium (\ion{Na}{1} -- \ion{Na}{3}), magnesium (\ion{Mg}{1} -- \ion{Mg}{4}), silicon (\ion{Si}{1} -- \ion{Si}{6}), sulfur (\ion{S}{1} -- \ion{S}{5}), calcium (\ion{Ca}{1} -- \ion{Ca}{5}), iron (\ion{Fe}{1} -- \ion{Fe}{5}), and electrons from an initial non-equilibrium state to a steady state. 

Modifications for solving the hydrodynamic equations include using an unsplit solver based on \citet{2013leeJCoPh.243..269L} as well as a hybrid Riemann solver that utilizes the Harten Lax and van Leer (HLL) solver \citep{einfeldt1991godunov} to ensure stability in regions of strong shocks, and the Harten-Lax-van Leer-Contact (HLLC) solver \citep{toro1994restoration,tororiemann} in continuous flows. MAIHEM was initially presented in \citet{gray2015atomic} and further improved upon in \citet{gray2016atomic} and \citet{gray2017effect}. 

The hydrodynamic equations solved by MAIHEM are given in the aforementioned papers and are invariant in space, time, and density under the transformation $x \rightarrow \lambda x,\ t \rightarrow \lambda t,\ \rho \rightarrow \rho/\lambda$, where $\lambda$ is an arbitrary number. Thus, the final steady-state abundances depend only on the mean density of the material in the simulation domain multiplied by the driving scale of turbulence, $n_{\rm tot}L$, the one-dimensional (1D) velocity dispersion of the gas, $\sigma_{\rm 1D}$, and the extragalactic UV background (EUVB) described by the ionization parameter $U$ such that,
\begin{equation}
\label{eq:1}
U  \equiv \frac{\Phi}{n_H c}, 
\end{equation}
where $\Phi$ is the total photon flux of ionizing photons, $n_{\rm H}$ is the hydrogen number density, and $c$ is the speed of light. We refer the reader to the original MAIHEM papers for further details.

The simulations are carried out in 128$^{3}$ periodic boxes, and they sample  metallicities of $Z/Z_{\odot}$ = 0.01, 0.1, 0.3, 1.0, 10. These simulations begin with a uniform density of $n_{\rm tot} = 5.0 \times 10^{-4}$ -- $2.5 \times 10^{1}$ cm$^{-3}$ and all have the box length set to $L_{\rm box} = 100$ pc on each side. Given this length, the driving scale of turbulence varies between $n_{\rm tot}L_{\rm box} = 1.5 \times 10^{17}$ and $7.6 \times 10^{21}$ cm$^{-2}$. Each run is initialized with a temperature of $10^5$~K and fractional ion abundances that correspond to collisional ionization equilibrium at this temperature.

Turbulence is driven with solenoidal modes \citep{pan2010} with wavenumbers varying in the range of $1 \leqslant L_{\rm box} \lvert k\rvert/2\pi \leqslant 3$, ensuring the average driving scale of turbulence is $k^{-1} \simeq 2L_{\rm box}/2\pi$. 
The  $\sigma_{\rm 1D}$ that we test is varied between $6$ and $60$~km~s$^{-1}$ ($\sigma_{\rm 3D}$ = 10 -- 100~km~s$^{-1}$) to obtain metal column densities in a wide range of turbulent conditions. 
The ultraviolet background for these runs is taken to be the redshift zero \citet{2012ApJ...746..125H} (HM2012) EUVB, the strength of which is allowed to incrementally increase in steps of $\approx 0.2$ between $-4 <$ log $U < -1$ following inferences made from observations done in the COS-Halos survey \citep{werk2014cos}. 

To determine when the box has reached a steady state, we monitor the average global abundances at every 10 time steps. To prevent ions with small abundances from stopping the progression of the ionization parameter, we consider the change in 
 fractional abundances, defined by
\begin{equation}
\frac{\Delta X_{\rm i}}{X_{\rm i}} = \frac{\overline{X_i^a} - \overline{X_i^b}}{\overline{X_i^a}},
\end{equation}
where $X_{\rm i}$ is the abundance of ion $i$ and $X_i^a$ and $X_i^b$ are the averaged ion abundances of the first 10 and last 10 time steps within an interval of 100 steps.
 During the MAIHEM simulations, when all fractional ion abundances are below a  cutoff value of 0.03, we record the ion fractions and move to higher $U$, until we reach the upper end of the grid at $\log U=-1$.

Once a simulation reaches a steady-state solution, we quantify the average ion fraction across the volume.  
With these values, we construct a grid of metal column densities with the resulting mass fractions from these MAIHEM models mapped onto \ion{H}{1} column densities ranging from $13.25 <$ log $N_{\rm H\ 1} < 22$ in steps of 0.15~dex.

By doing so in models of different volume density, we are effectively generating column densities on path-lengths that are not limited to the size of the simulation volume, taking the simulation only as a representative patch over which we can compute the average ion fractions. However, this approximation breaks down for scales that are much smaller than the driving scale of the turbulence, for which our patch is not representative anymore. We control for this assumption explicitly in our MCMC analysis by means of a dedicated prior, as described below.

We further note that the simulations are run in the optically-thin limit, and hence our modelling will start becoming inaccurate in the high-column density regime where gas will start to self-shield. As shown below, however, the majority of the systems can be found at $\log N_{\rm H\ I}\lesssim 10^{19}~\rm cm^{-2}$, a regime where some radiative transfer effects start to become relevant but where the gas is still highly ionized.

The final grid of models, with all parameters and their ranges, is summarized in Table \ref{tab:table1}. It should be noted that we 
allow for a redshift parameter due to the current implementation of the MCMC procedure which takes redshift as an input, but in fact we assume the same mass fractions across the range in redshift. The (modest) variation of the UVB across this redshift range is however still captured by the varying ionization parameter.  

It is also worth noting that 
the final quantity that we compare against observations is the mass fraction of ions averaged over a simulated box. This is the most appropriate description of a turbulent medium in which pockets of gas of different density and temperature coexist in a steady state. By doing so, however, we are modeling exclusively systems in which components of different ionization state arise from the same medium, and not systems in which multiple components arise from different gas patches superimposed in velocity along the line of sight. For this reason, we expect the MAIHEM model to work best to describe the co-spatial case in which components are preferentially aligned in velocity space. Conversely, for systems that do not preferentially show co-spatial components, our model is likely to not capture the real complexity of multiple gas phases that are superposed in projection. In this case, we expect not to perform any better than the case of PIE models, which make the same simplifying assumption.

\begin{table}[]
\centering
\begin{tabular}{ccc}
\hline
\hline
Parameter & Min. & Max. \\
\hline
log $N_{\rm H\ I}\ (\rm cm^{-2}$)  & 13.25 & 22.00  \\
z  & 0.0  & 0.4 \\
log $Z/Z_{\odot}$ & -2  & 1  \\
log U & -4  & -1 \\
$\sigma_{\rm 3D}\ (\rm km~s^{-1}$)& 10   & 100 \\
log $N_{\rm tot}\ (\rm cm^{-2}$) & 17.26   & 21.96 \\
\hline
\end{tabular}
\caption{Summary of parameters and their range for the MAIHEM model grid.}
\label{tab:table1}
\end{table}

\subsection{Emcee Procedure}
To directly compare the MAIHEM models to observations and infer underlying physical properties from the data, we use a Bayesian approach where the likelihood is computed using the affine invariant MCMC sampler  {\sc emcee} \citep{foreman2013emcee}.
Our analysis closely follows previous work
based on PIE models \citep[][P17]{2015crighton,fumagalli2015physical}, where we replace the grid of models with the one computed from the MAIHEM simulations.

The basic idea is to compute the posterior probability distribution function (PDF) for a set of observations, $N$, given a model, $M$, and set of parameters, $\Theta$. This may be represented functionally as
\begin{equation}
P(\Theta|N,M)=\frac{p(\Theta|M) L(N|\Theta,M)}{Z},
\end{equation}
where $p(\Theta|M)$ is a prior to constrain the allowed parameter space, $L(N|\Theta,M)$ is the likelihood function, and $Z$ is the normalization. 
In the context of this analysis, we use {\sc emcee} \citep{foreman2013emcee} to constrain the posterior distribution of parameters of interest (e.g. metallicity and turbulence velocity) given observed column densities of different ions for a given system, and the corresponding modelled column density. 

More specifically, our likelihood function is defined by
\begin{equation}
L(N|\Theta) = \prod_{\rm i}\frac{1}{\sqrt{2\pi} \sigma_i} \exp\Big(-\frac{(N_i-\overline{N_i}(\Theta))^2}{2\sigma_i^2}\Big),
\end{equation}
where $N_i$ is the observed column density for the $i$-th ion, $\overline{N_i}(\Theta)$ is the model column density for the same ion, and $\sigma_i$ is some error associated with the observed column density. The product is taken over all the ions for one system. 
As in \citet{fumagalli2015physical}, the likelihood is modified in presence of lower/upper limits on the ions, to include the product of the Gaussian integral over all possible column densities allowed by the observations. We then take the median of the resulting posterior as the best estimate for parameters of interest.

\begin{figure}[t]
     \hspace*{-0.3cm}
     \vspace*{-0.5cm}
    \includegraphics[width=1.08\linewidth]{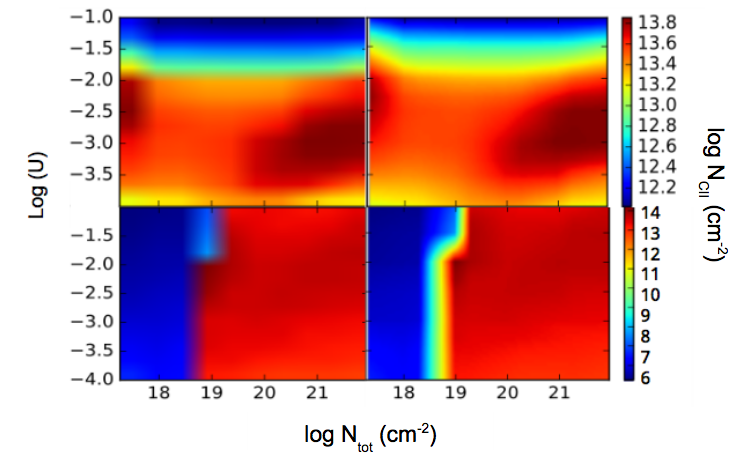}
    \caption{\ion{C}{2} column density slices in the Log $U$ versus log $N_{\rm tot}$ plane, shown through the grid at log $N_{\rm H\ I}=16.25$ cm$^{-2}$ and $[Z/Z_{\odot}]=0.3$. The top panels show a case of low turbulence, with $\sigma_{\rm 3D}=10$ km s$^{-1}$, while the bottom panels show a case of high turbulence, with $\sigma_{\rm 3D}=80$ km s$^{-1}$. The simulated log $N_{\rm C\ II}$ is shown on the left, while the column density interpolated on a finer grid is shown on the right.}
    \label{fig:c2_slices}
\end{figure}

Our prior conditions are informed by the MAIHEM models and are given here. As described above, we take mass fractions from the MAIHEM models and map them onto a range of \ion{H}{1} columns, thus using the simulation boxes as representative patches. To avoid solutions where the physical size of the absorber is below the scale at which we drive turbulence, we implement a physical versus non-physical prior such that solutions are constrained to $N_{\rm H\ I} \geqslant n_{\rm H} x_{\rm H\ I} L$. When evaluating samples within the grid, we take Gaussian priors on the redshift and on the measured $N_{\rm H\ I}$, with width equal to the observational error. In cases where the $N_{\rm H\ I}$ is constrained by limits (e.g. for saturated Lyman series lines), we use instead a top-hat function that brackets these limits. Flat priors spanning the entire grid of models are instead assumed for the remaining parameters. 

During our analysis, we use 100 walkers with 300 samples and a burn-in phase of 150 steps. We initialize the $N_{\rm H\ I}$ and redshift at the observational value. We then slightly offset each walker in the following way,
\begin{equation}
    pos = a\times (2\times\sigma_{\rm H\ I})+(N_{\rm H\ I}-\sigma_{\rm H\ I}),
\end{equation}
where $a$ is a random number drawn from a uniform distribution between 0 and 1 and $\sigma_{\rm H\ I}$ is the error on the $N_{\rm H\ I}$ measurement. All the other axes are initialized randomly across the full range of that axis. Throughout our analysis, we find that some of the fits are noisy and thus we re-run the MCMC procedure initializing the walkers at the median solution of the previous noisy run. This improved the acceptance fraction as well as reduced the frequency of walkers stuck in a bad region of parameter space.
Finally, to ensure that the MCMC has fully converged with our number of walkers and samples, we conduct additional test runs on a selected number of systems. These runs use 300 walkers for 600 samples, but find the same solutions as in the case of 100 walkers with 300 samples. 

At each step, the MCMC procedure evaluates the likelihood given above, and we use the {\sc SciPy} regular grid linear interpolator to evaluate the likelihood in regions of parameter space that were not simulated in between the grid points. Through the above priors, we restrict the interpolation to within the grid, not allowing for extrapolations outside the  grid domain. It is worth noting that, especially for the case of high-turbulence, the model column density presents some regions of steep gradient across parameter space.  

An example is shown in 
Figure \ref{fig:c2_slices} for the \ion{C}{2} column density in a case of low (top left; $\sigma_{\rm 3D} = 10~km~s^{-1}$) and high (bottom left; $\sigma_{\rm 3D} = 80~km~s^{-1}$) turbulence. Care has been taken to ensure that the model grid has sufficient resolution to allow for a sensible interpolation across regions of steep gradient. As shown in Figure~\ref{fig:c2_slices}, once we interpolate the original grid on finer steps (compare left and right panels), we recover the same features present in the original model although, as expected, the width of the discontinuity is slightly broadened by the interpolation scheme.  

\subsection{Application to COS-Halos data}

Before applying the MCMC analysis to the observations, it is instructive to consider how  MAIHEM models differ from the pure case of PIE.
To this end, we can turn again to the example presented in Figure~\ref{fig:c2_slices}. \citet{gray2015atomic,gray2016atomic,buie2018modeling} showed that at low amounts of turbulence, $\sigma_{\rm 3D} \leqslant 10$ km s$^{-1}$ (top panel), MAIHEM mass fractions agree well with the pure photoionization case computed with CLOUDY \citep{ferland2013}.

\begin{figure}[h]
    \centering
    \includegraphics[width=1.08\linewidth]{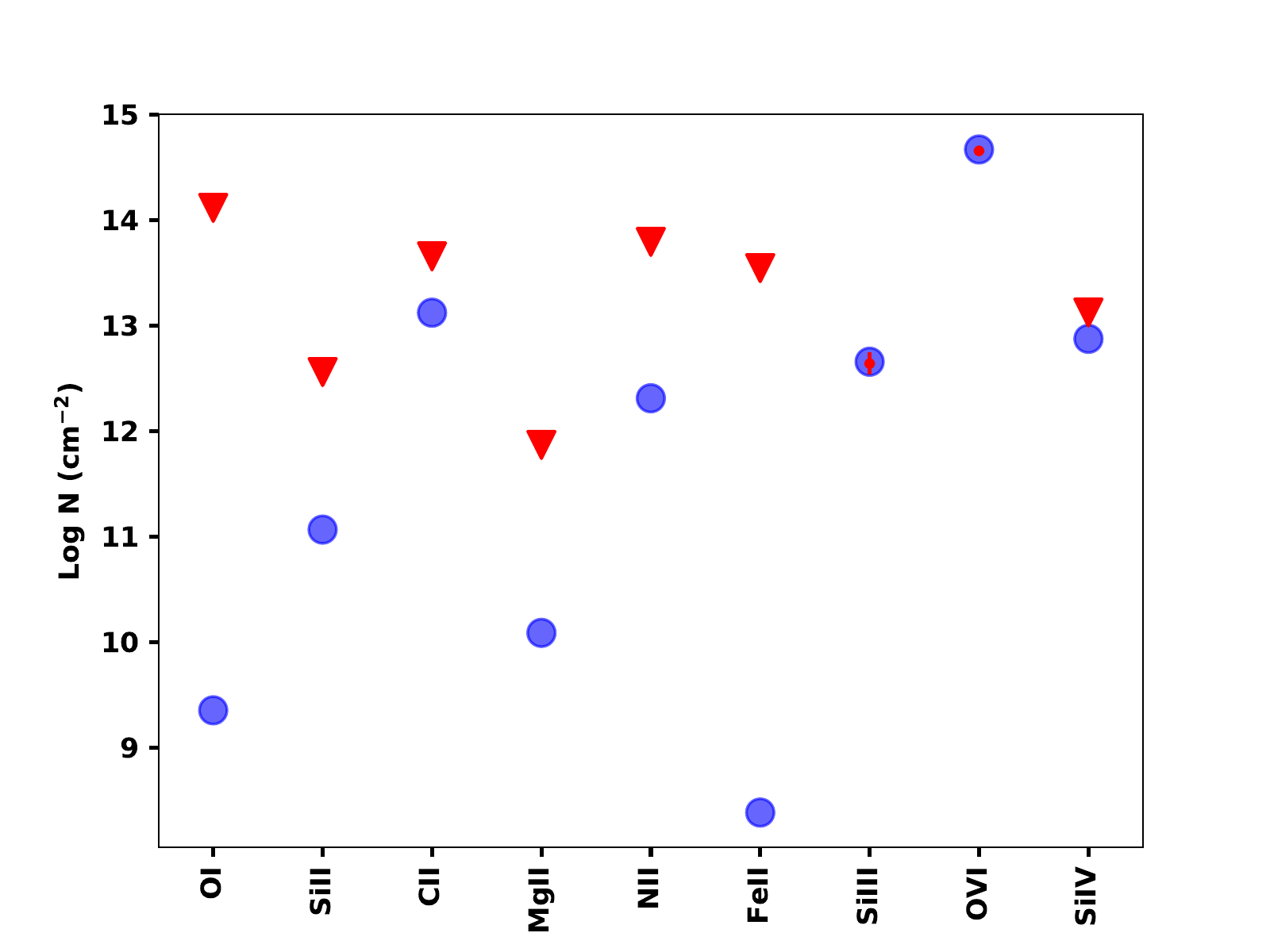}
    \includegraphics[width=1.08\linewidth]{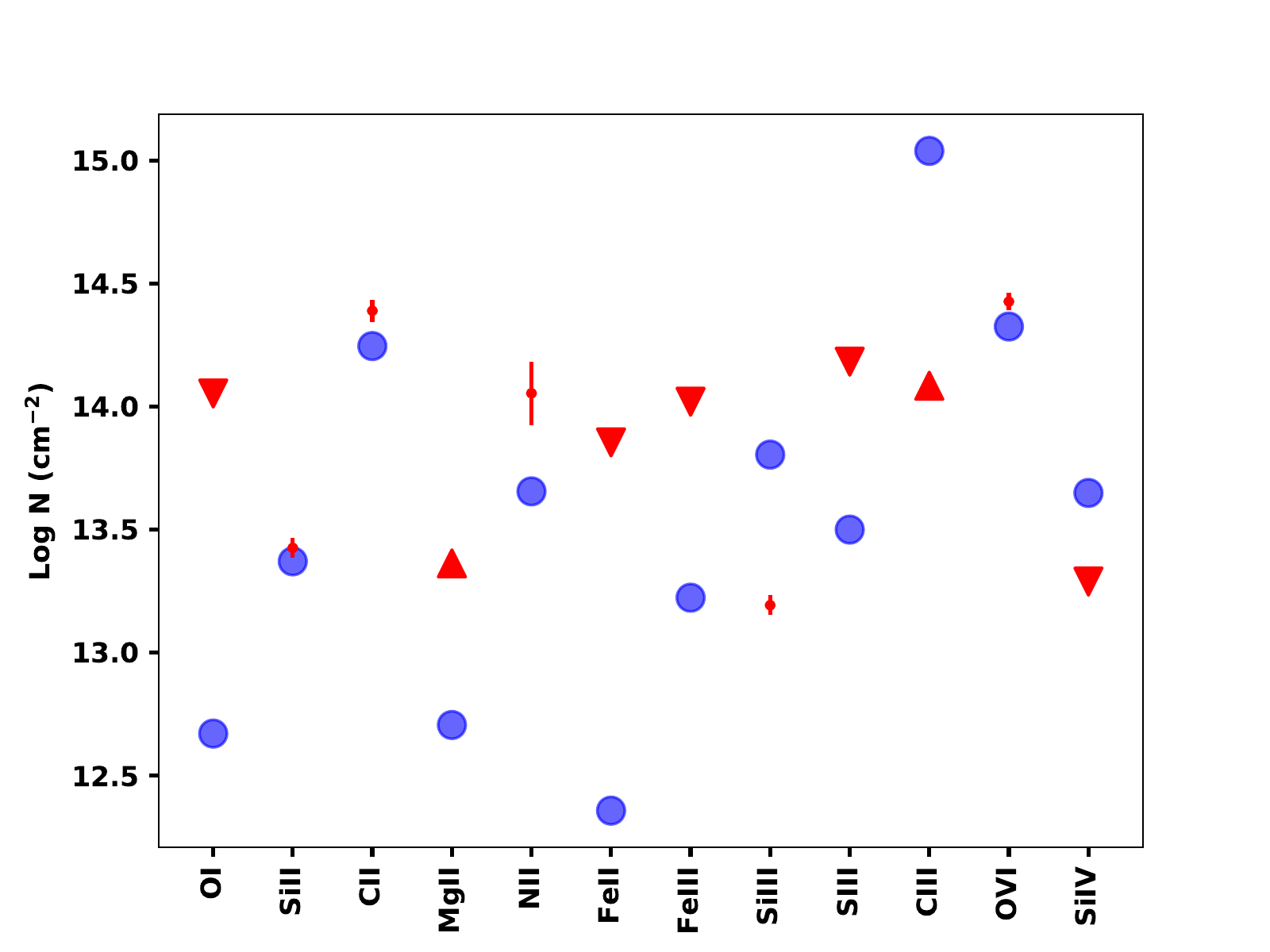}
    \caption{Ion by ion comparison between observations and models for the systems J0914+2823\_41\_27 and J1330+2813\_289\_28. We show two representative cases, one in which the observations are well reproduced by the models (top) and one in which some of the ions (e.g. \ion{N}{2} and \ion{Si}{3}) show some discrepancies (bottom). The observed column densities are in red (arrows are limits while points indicate a measurement with an associated error) and the model best fit column densities are in blue.}
    \label{fig:residuals}
\end{figure}

\begin{figure*}[h]
    \centering
    \includegraphics[width=1.0\linewidth,height=1\textheight]{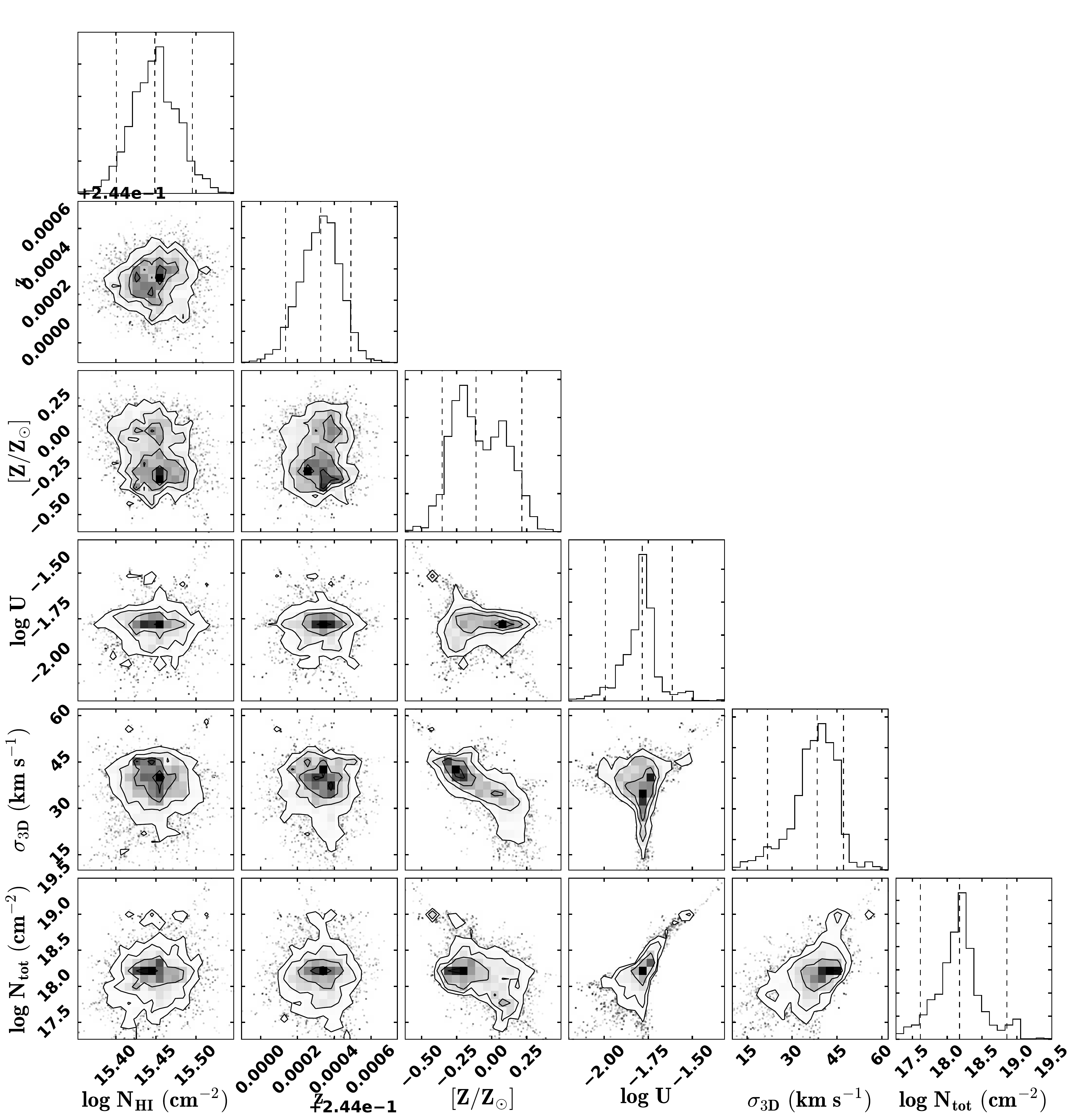}
    \caption{A corner plot for system J0914+2823\_41\_27, showing the distribution of samples over the parameter space.}
    \label{fig:sys_good_fit1}
\end{figure*}

\begin{figure*}[h]
    \centering
    \includegraphics[width=1.0\linewidth,height=1\textheight]{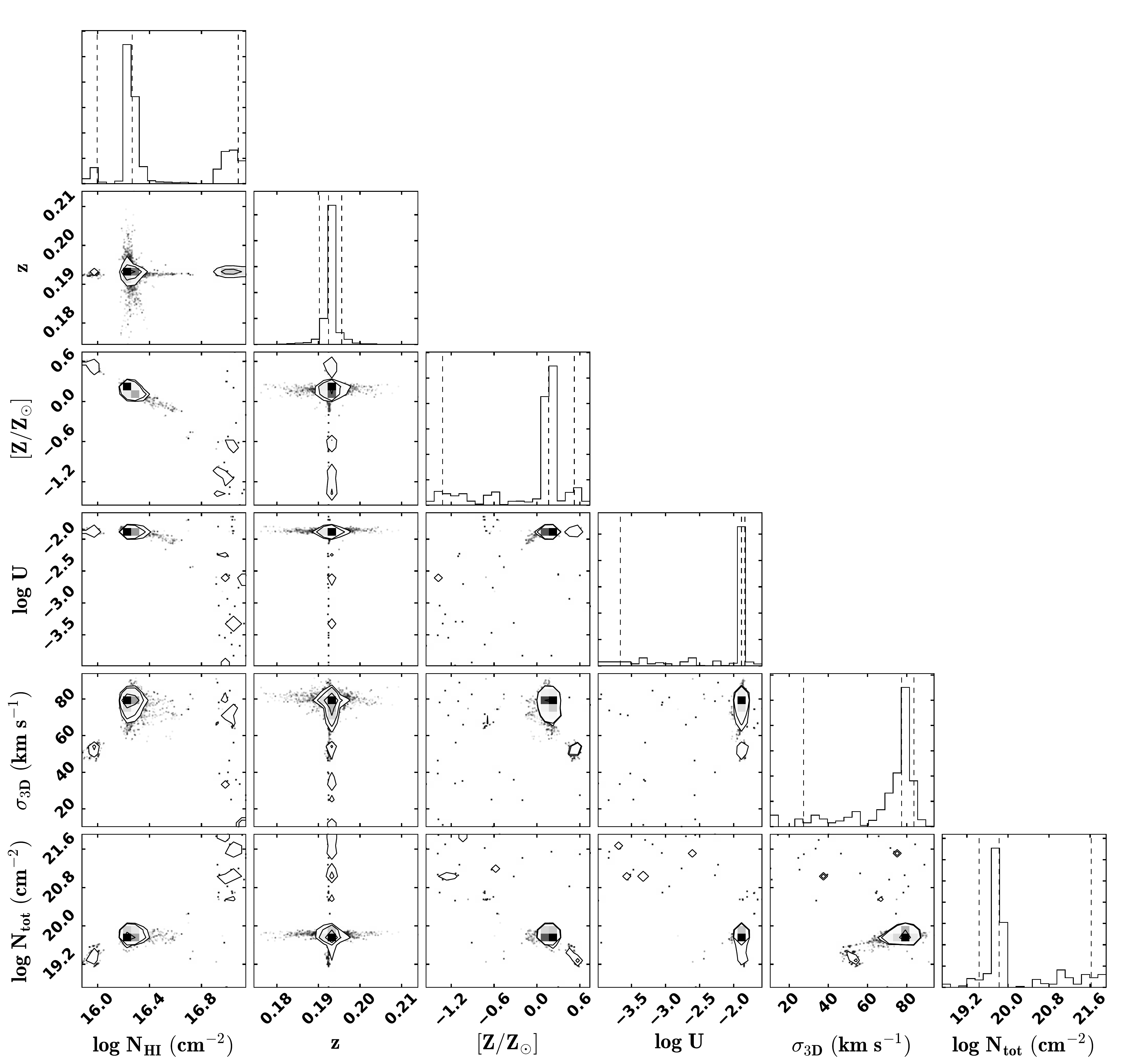}
    \caption{Same as Figure~\ref{fig:sys_good_fit1}, but for J1330+2813\_289\_28.}
    \label{fig:sys_good_fit2}
\end{figure*}

We see that merely changing the turbulent velocity results in a stark difference in the distribution of low ions, like $N_{\rm C\ II}$. In a purely photoionized medium, the simulation domain is dominated by a single ionization state that traces the equilibrium temperature of the gas. This results in a nearly smooth distribution of $N_{\rm C\ II}$ across $N_{\rm tot}$. There is however a sharp boundary at log$(U) \gtrsim -2$, where this ion rapidly drops off in abundance as it is ionized to the intermediate ions \ion{C}{3} and \ion{C}{4}. 

This behaviour changes instead when the turbulent velocity is increased to $\sigma_{\rm 3D}=80$ km s$^{-1}$ as \ion{C}{2} is nearly absent for $N_{\rm tot} \lesssim 10^{18.5}$ cm$^{-2}$, followed by a rapid increase in abundance. As the turbulent velocity is increased, the simulation domain temperature increases, such that total columns below a certain amount do not have enough metals to provide more cooling, thus finding equilibrium at points that are dominated by intermediate and high ions with small clusters of cold gas housing low ions. 

We are now ready to apply our procedure to the full COS-Halos sample. When comparing models and observations, we treat absorbers as total systems rather than independent clouds, in line with the definition of \citet{prochaska2015}. This choice is not unique \citep[see e.g.][]{2016ApJ...833..283L,2019arXiv191004310K}, but as argued above, it is the most appropriate for describing a turbulent medium in which small pockets or material at different temperatures and densities are changing rapidly but are statistically described by averages once a steady state is reached.

Results for the posterior probability distribution are summarized in Table~\ref{tab:tableall}. We also present in Figure~\ref{fig:residuals} two examples of the results through a ion by ion comparison of observations and best fit models for the systems J0925+4004\_196\_22\footnote{Throughout this analysis, we follow the nomenclature of P17 to label systems.} and J1330+2813\_289\_28. In both cases, the observed ion column density is shown as either a red arrow indicating a limit or a red point with error bars. The best fit model column density computed from the grid at the median values of the posterior PDFs is in blue. The associated corner plots for these systems are shown in Figures \ref{fig:sys_good_fit1} and \ref{fig:sys_good_fit2}.

We specifically choose to show a system with high level of agreement between the model fits and the observed columns (J0925+4004\_196\_22), and one that finds less agreement between the two (J1330+2813\_289\_28). Looking at the residual plot for our high-agreement example, we see that the model is generally a fairly accurate description of the data, with the best fit model column densities either agreeing with limits or overlapping with the observations. It should be noted how the MAIHEM model is able to describe at the same time low-ions and \ion{O}{6}. J1330+2813\_289\_28, however, shows a more mixed agreement, with a subset of ions agreeing well with observational constraints, while others showing some level of discrepancy. Fits of this quality are also encountered when using PIE models, and represent a small subset of the entire sample, as shown below more quantitatively. In general, the corner plots show symmetrical ellipses for the posterior PDFs, however, there do appear to be some correlations among parameters, such as between both $N_{\rm tot}$ and $\sigma_{\rm 1D}$ and the metallicity as well as $U$ and $N_{\rm tot}$.

\begin{figure*}[t]
\centering
\hspace*{-0.4in}
    \includegraphics[width=1.1\linewidth]{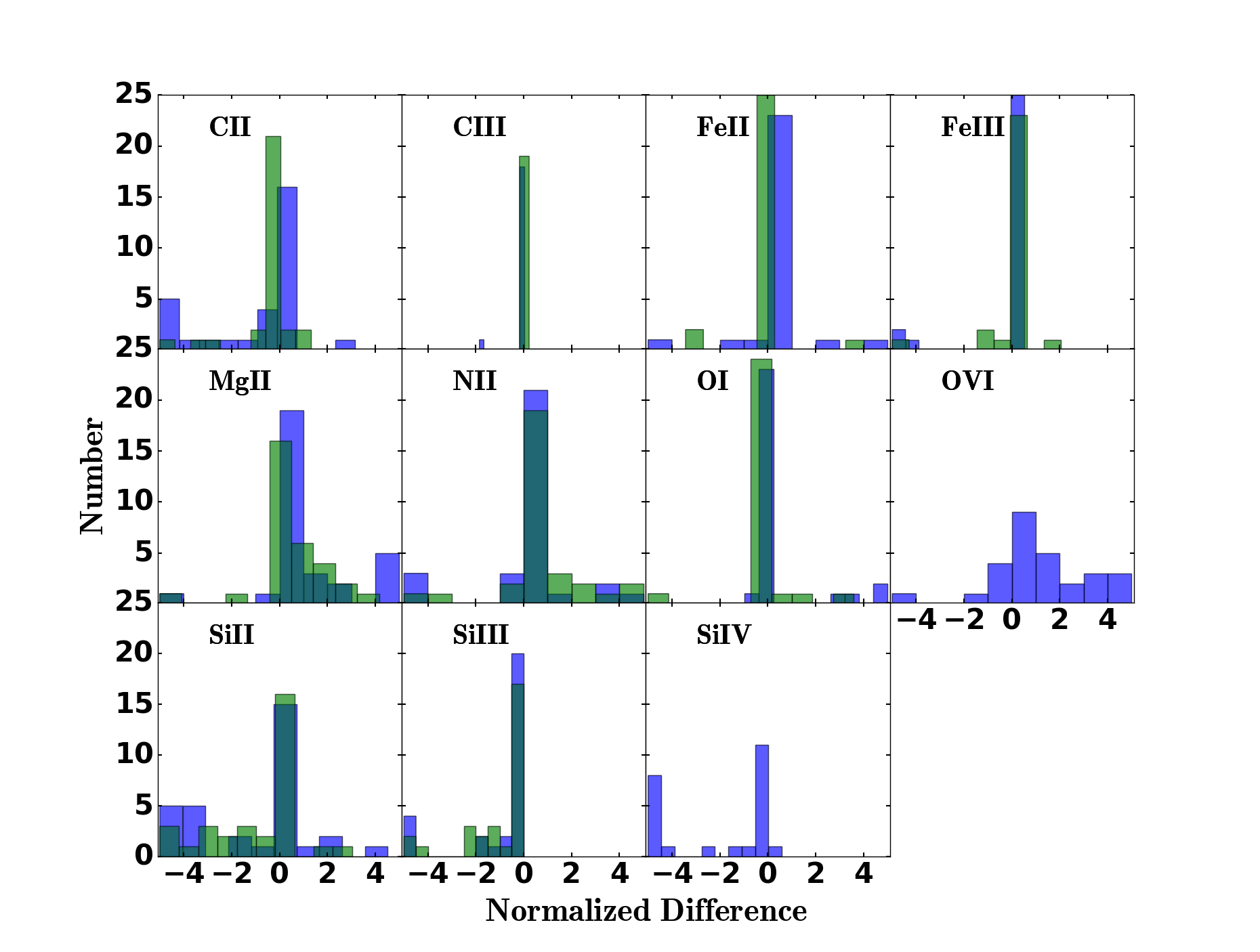}
    \caption{Histograms showing the distributions of differences between the observed column densities and the best-fit model column densities normalized by the observational error for MAIHEM (blue) and PIE (green). There are a few cases of these differences being larger than $\pm 5$, but we show them  at $\pm 5$.}
    \label{fig:ion_hist}
\end{figure*}

\section{Results}

\subsection{Metal Ion Fits}

With all the posterior PDFs in hand, we can investigate more systematically how well the MAIHEM model describes the data in comparison to PIE models, and investigate what we can learn from the derived posterior PDFs. 

We begin by quantifying how well ions are reproduced by our model, by quantifying the difference between the observed column density and the modeled column density normalized by the error in the observations,
\begin{equation}
    \Delta N/N = \frac{N_x-N(\Theta)_x}{\sigma_x},
\end{equation}
where $N_{\rm x}$ is the observation column, $N(\Theta)_{\rm x}$ is the best fit modeled column (evaluated at the median of the posterior PDFs), and $\sigma_{\rm x}$ is the error in $N_{\rm x}$. Here, positive (negative) differences indicate that the model under-predicts (over-predicts) the observed column density. We deal with limits by asserting no difference if the fit is on the correct side of the limit and a normal difference otherwise, where the error in the normalization of the limit is taken to be the average error in column densities for a system. This is done for \ion{C}{2}, \ion{C}{3}, \ion{Fe}{2}, \ion{Fe}{3}, \ion{Mg}{2}, \ion{N}{2}, \ion{O}{1}, \ion{O}{6}, \ion{Si}{2}, \ion{Si}{3}, and \ion{Si}{4} and shown in Figure \ref{fig:ion_hist} ( blue histograms). \ion{S}{3} is also included in our analysis, but we do not plot its difference as there are only 5 systems with this ion, 4 of which are limits, and they all have no difference. Also, there are a few cases of these differences being larger than $\pm 5$, and for visualization purposes we plot them at $\pm 5$.

In general, for most of the systems, we are able to find a model solution that agrees well with the observed column densities. Considering the tail of models with more deviant solutions, we see that \ion{C}{2}, \ion{C}{3} (although this is only one observation out of 19), \ion{Si}{3}, and \ion{Si}{4} tend towards solutions whose model column densities are higher than the observed ones (both Fe ions and \ion{N}{2} show this as well, however only slightly). In contrast \ion{Mg}{2}, \ion{O}{1}, and \ion{O}{6} show the opposite trend. \ion{O}{6}, in particular, exhibits a fairly Gaussian distribution of normalized differences around zero with a tail extending in both directions, although favoring positive differences. 

A key result of this analysis is that turbulence
as included in these models promotes the density and temperature gradients needed to sustain both low, intermediate, and high ions simultaneously \citep{buie2018modeling}. We see, however, that the MAIHEM model does not provide a perfect description of the data for  intermediate and high ions. For this comparison, we have focused on the model prediction evaluated at the median of the posterior PDFs, which show a broad range of possible values. Thus, acceptable model predictions span a range beyond the single value adopted here and this is likely to ease some of the difference between models and observations. At the same time, the MAIHEM model is built under a simplifying set of assumptions, and therefore it is not expected to be a full description of the turbulence that is expected within the CGM. 

In Figure \ref{fig:ion_hist}, we also show the normalized difference between PIE models from P17 and observations (over-plotted in green). Furthermore, we quantify the accuracy of each ion fit by finding the percentage of normalized differences that lie within $\pm 1\sigma_{\rm obs}$ of $N_{\rm obs}$, as shown in Table \ref{tab:table2}. As above, for limits, we take the average error of column densities belonging to a system and apply it to the the limits when finding their normalized differences. We have one system, J1619+3342\_113\_40, which only had columns with limits in P17, while presently this system has non-limit measurements for \ion{Si}{4} and \ion{O}{6}. Given this, we find the average error for those ions and use that values for the remainder of the column densities for this system. 

We find that some ions are more accurately modelled by PIE (e.g. \ion{C}{2}) while others are more accurately described by the MAIHEM models rather than the PIE counterparts (e.g. \ion{N}{2}). However, in general, the two models appear to  yield fairly close results with the accuracy of ion fits being different by more than 10\% only in the case of \ion{C}{2}. Also, similarly to the MAIHEM model, the PIE model finds solutions that over-predict the column densities for \ion{C}{2}, \ion{Si}{2}, and \ion{Si}{3} and under-predicts them for \ion{Mg}{2} and \ion{O}{1}. 
Altogether, this comparison shows that the MAIHEM model is able to provide a description of the low and intermediate ions that is comparable to the one offered by PIE, with the added value of capturing at the same time the multiphase nature of the CGM needed to describe high ions.

\begin{table}[]
\centering
\begin{tabular}{lccc}
\hline
\hline
Ion         & MAIHEM & PIE$^{a}$ & Number$^{b}$  \\
\hline
\ion{C}{2}  & 67\%   & 83\% & 30 \\
\ion{C}{3}  & 95\%   & 100\% & 19 \\
\ion{Fe}{2} & 86\%   & 89\% & 28 \\
\ion{Fe}{3} & 89\%   & 89\% & 28 \\
\ion{Mg}{2} & 65\%   & 58\% & 31 \\
\ion{N}{2}  & 77\%   & 68\% & 31 \\
\ion{O}{1}  & 86\%   & 89\% & 28\\
\ion{O}{6}  & 46\%   & --  & 28 \\
\ion{Si}{2} & 50\%   & 56\% & 32 \\
\ion{Si}{3} & 76\%   & 70\% & 30 \\
\ion{Si}{4} & 54\%   & -- & 24  \\
\hline
\end{tabular}
\caption{Comparison of the performance of MAIHEM and PIE models. $^{a}$ \ion{Si}{4} and \ion{O}{6} are not included in the PIE analysis. $^{b}$ Number of observations for each ion.}
\label{tab:table2}
\end{table}

\subsection{Properties of a Turbulent CGM}

\begin{figure}[h]
    \includegraphics[width=1.0\linewidth]{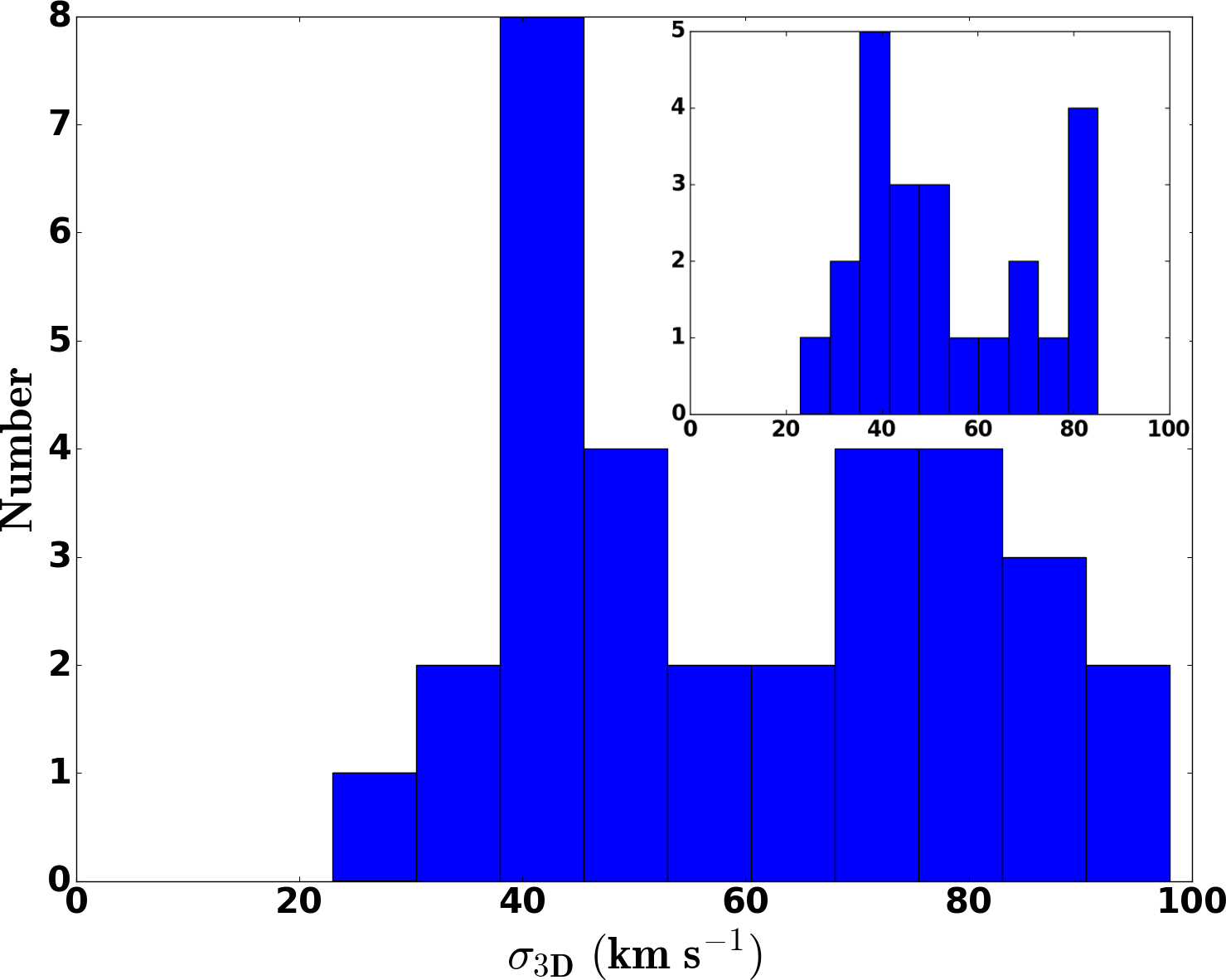}
    \caption{Histogram showing the median turbulent velocity distribution. We include a sub-histogram for systems with at least 67\% of their best fit model columns being within $\pm 1\sigma_{\rm obs}$ of $N_{\rm obs}$. Axis labels are the same as the larger histogram.}
    \label{fig:turb_hist}
\end{figure}

All of the systems converge on solutions that have some amount of turbulence, with the lowest being $\sigma_{\rm 3D} = 23$ km s$^{-1}$ for the system J2345-0059\_356\_12. We show a histogram of the median PDFs of the turbulent velocities in Figure \ref{fig:turb_hist}, where we see two peaks in the $\sigma_{\rm 3D}$ values around 40 and 75 km s$^{-1}$, with a tail extending to higher turbulent velocities. 
To further investigate if there is a preferred turbulent velocity among systems for which MAIHEM models provide an accurate fit, we focus on systems with at least 67\% of their best fit model columns being within $\pm 1\sigma_{\rm obs}$ of $N_{\rm obs}$ (shown in the inset). For this subset, we clearly see again a peak around 40~km~s$^{-1}$, while the second peak is less clear due to small number statistics. A tail beyond 60~km~s$^{-1}$ persists in this case.
This finding reinforces the idea that turbulence is required to model the CGM, in line with the observations of the line widths \citep{werk2016ApJ...833...54W}.

\begin{figure}[b]
    \includegraphics[width=1.0\linewidth]{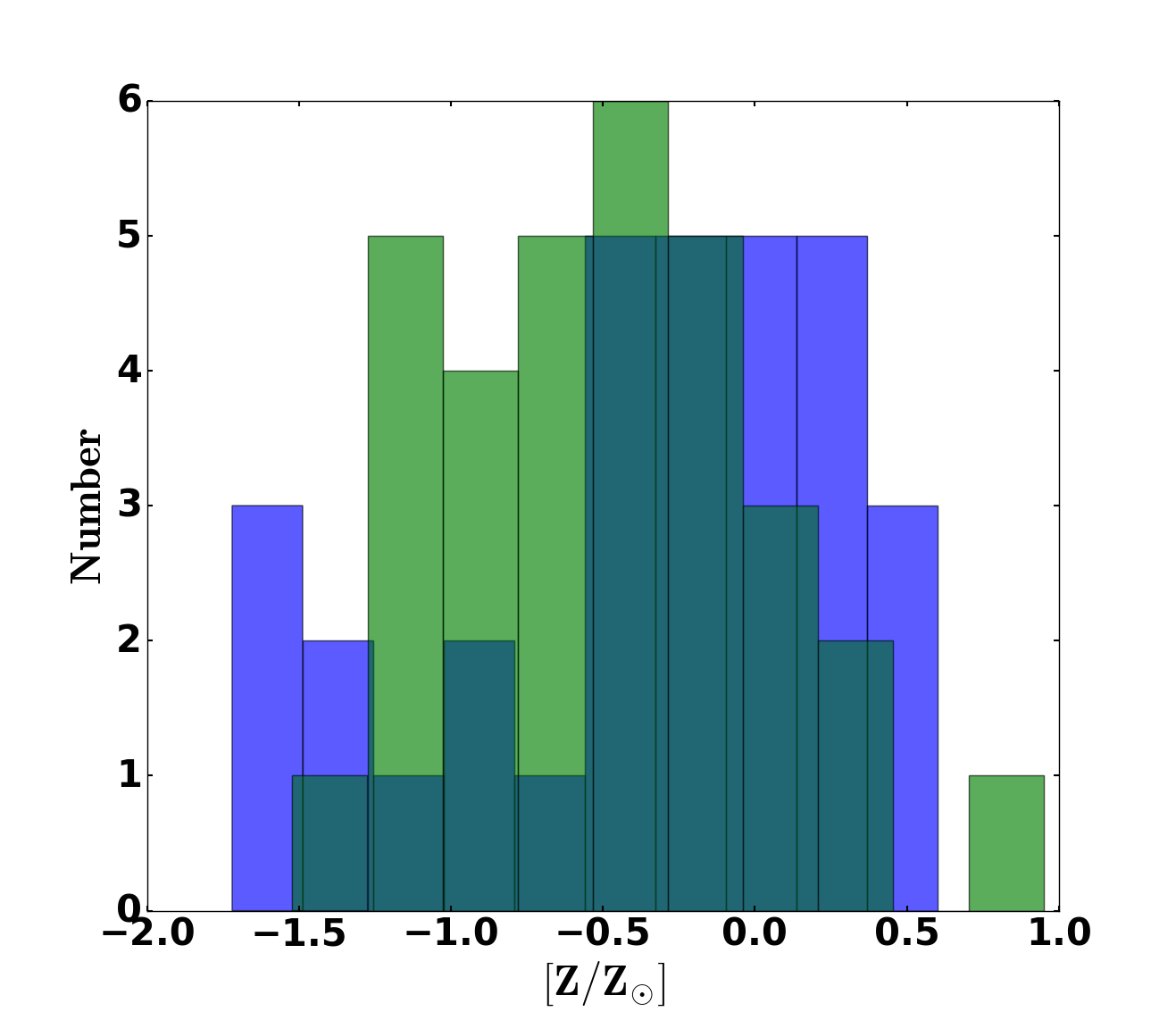}
    \caption{Histogram showing the median metallicity distribution inferred using the MAIHEM (blue) and PIE (green) models.}
    \label{fig:met_hist}
\end{figure}

Having assessed the posterior PDF of the turbulence velocity, we examine next results for the median metallicity inferred using MAIHEM models, comparing again with the results of P17. This is shown in Figure \ref{fig:met_hist}, where the results for MAIHEM are in blue while results from P17 are in green. We see that the two histograms overlap with one another, however MAIHEM seems to prefer solutions in which the gas is  more enriched when compared to P17, with typical values approaching solar metallicity. Although it is not trivial to disentangle the origin of this trend, 
we speculate that the extra turbulent energy input adds heat to the gas, leading to solutions at higher metallicities which provide higher cooling rates. At the same time, the inclusion of \ion{O}{6}, which  is likely to trace a hot and enriched phase \citep{2014ApJ...788..119L}, may contribute to skew the solution towards higher metallicity. Regardless to the physical origin, this analysis shows that key inferred quantities for the CGM, such as metallicity, are somewhat dependent on the model adopted. 

\begin{figure}[t]
    \includegraphics[width=1.08\linewidth]{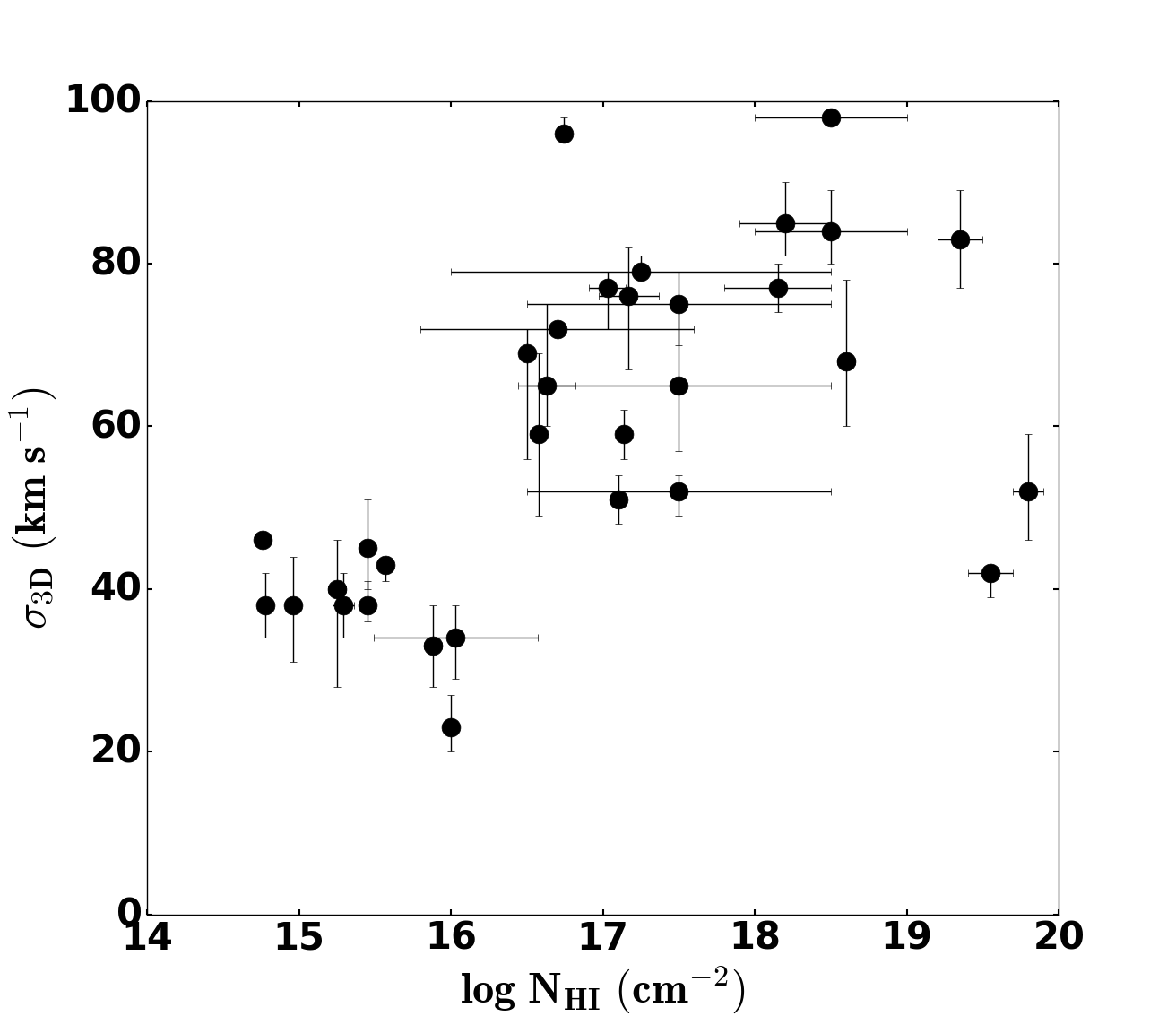}
    \caption{Scatter plot of the median inferred $\sigma_{\rm 3D}$ as a function of the observed $N_{\rm H\ I}$. Error bars for $\sigma_{\rm 3D}$ represents the probability contained between the 34th (lower limit) and the 68th (upper limit) percentile.}
    \label{fig:H1vsturb}
\end{figure}

Next, we investigate on a system-by-system basis what trends exist among the inferred physical parameters and observations, to learn about the turbulent nature of the CGM. We first show the inferred $\sigma_{\rm 3D}$ for individual systems as a function of the observed $N_{\rm H\ I}$ in Figure \ref{fig:H1vsturb}. We find a positive correlation with $>99.9\%$ statistical significance. 

\begin{figure}[h]
    \includegraphics[width=1.08\linewidth]{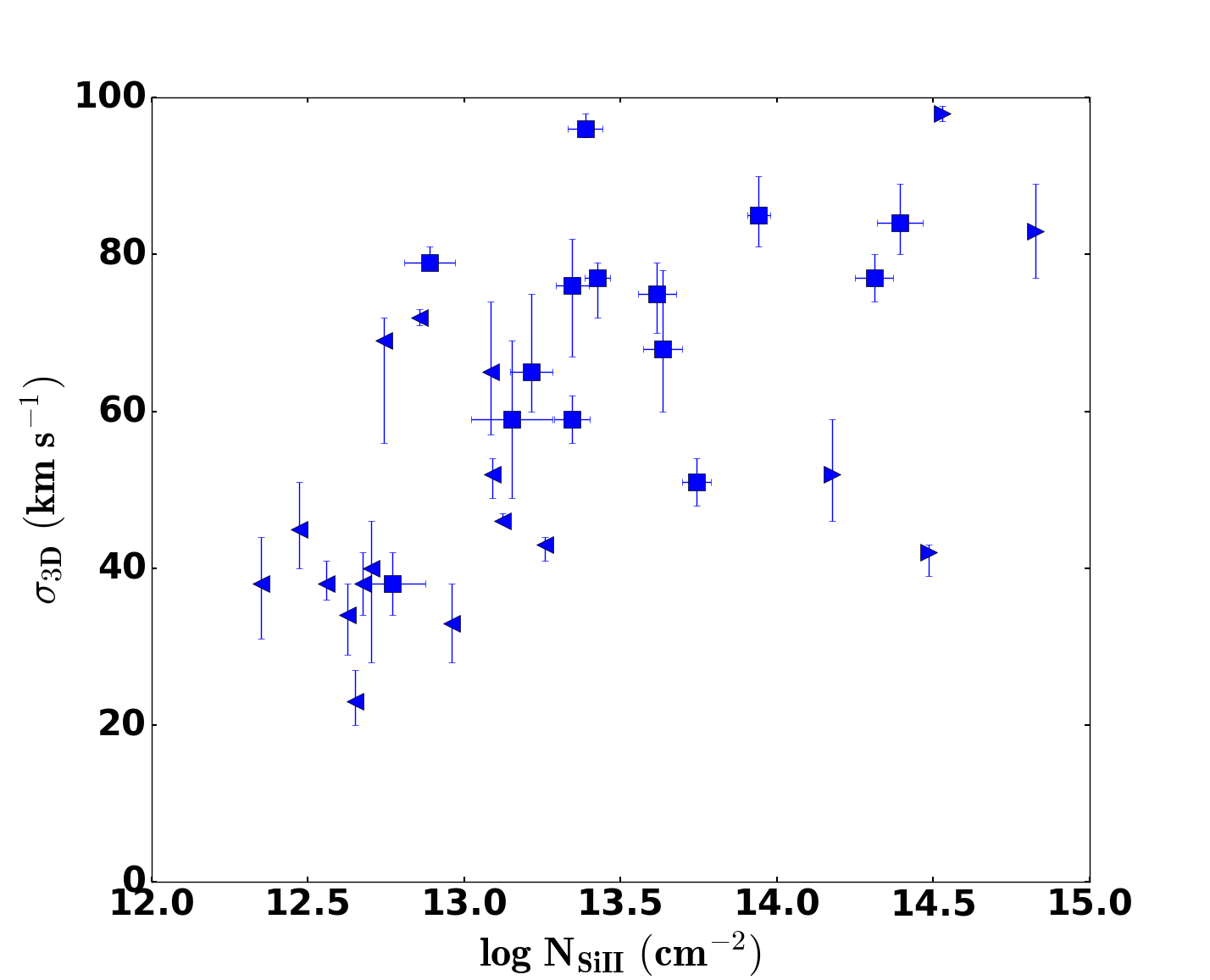}
    \caption{Scatter plot of the median inferred $\sigma_{\rm 3D}$ as a function of the observed $N_{\rm Si\ II}$. Error bars for $\sigma_{\rm 3D}$ represents the probability contained between the 34th (lower limit) and the 68th (upper limit) percentile.}
    \label{fig:SiIIvsturb}
\end{figure}

We also investigate trends with low ions such as \ion{Si}{2}, which tend to populate the inner region of galactic halos \citep[see e.g. figure 4][]{2013werkApJS..204...17W}. We plot the inferred $\sigma_{\rm 3D}$ as a function of the
observed $N_{\rm Si\ II}$  in Figure \ref{fig:SiIIvsturb}, finding again a positive correlation at the $>99.9\%$ confidence level. It should however be noted that 16/32 $N_{\rm Si\ II}$ are lower or upper limits, with the lower limits being more prevalent at lower turbulent velocities and upper limits at higher values of turbulence. Given these limits, this correlation may be skewed even more in the positive direction.

\begin{figure}[t]
    \includegraphics[width=1.08\linewidth]{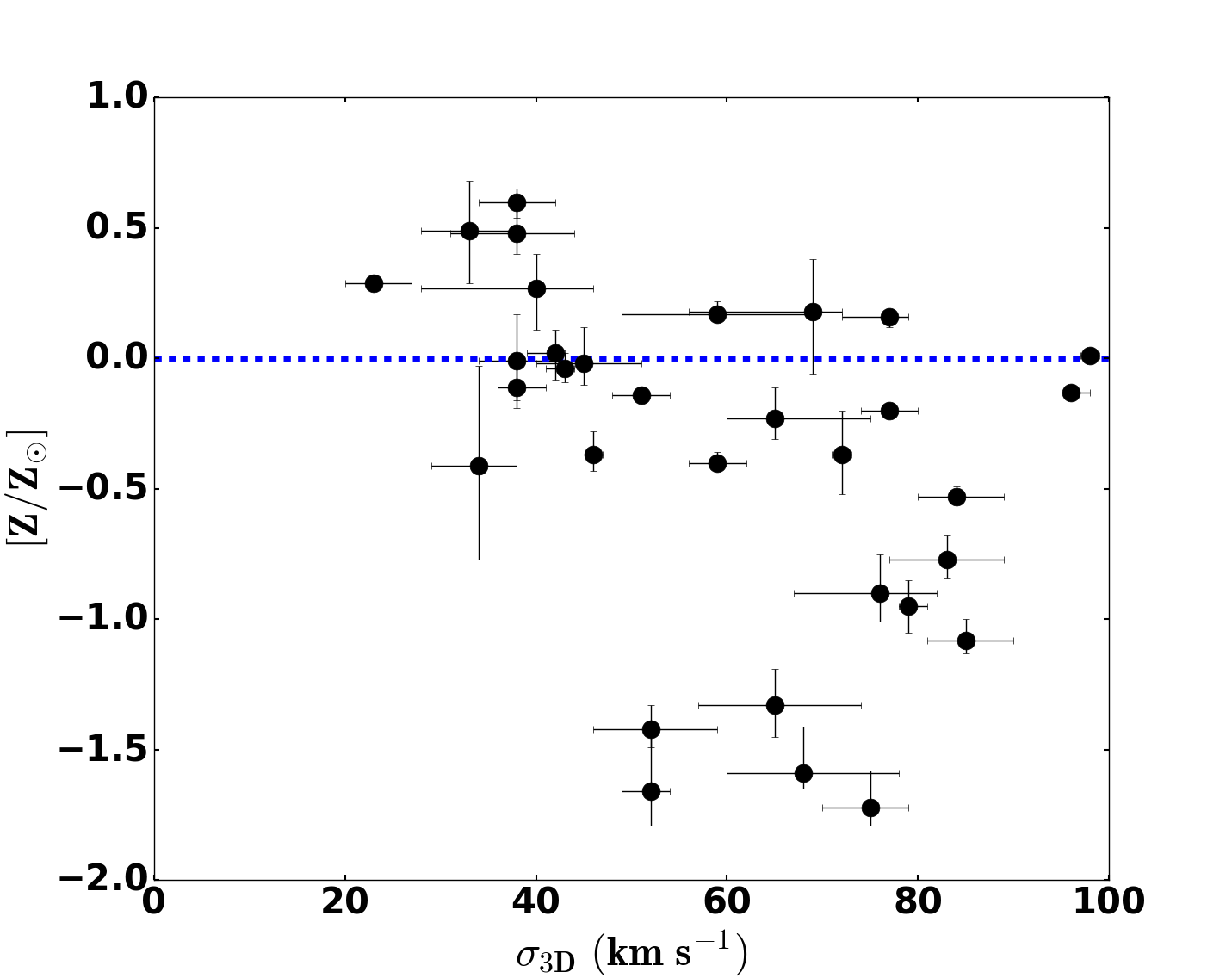}
    \caption{Scatter plot of the median inferred metallicity as a function of the median $\sigma_{\rm 3D}$. Error bars represents the probability contained between the 34th (lower limit) and the 68th (upper limit) percentile.}
    \label{fig:metvturb}
\end{figure}

\begin{figure}[t]
    \includegraphics[width=1.08\linewidth]{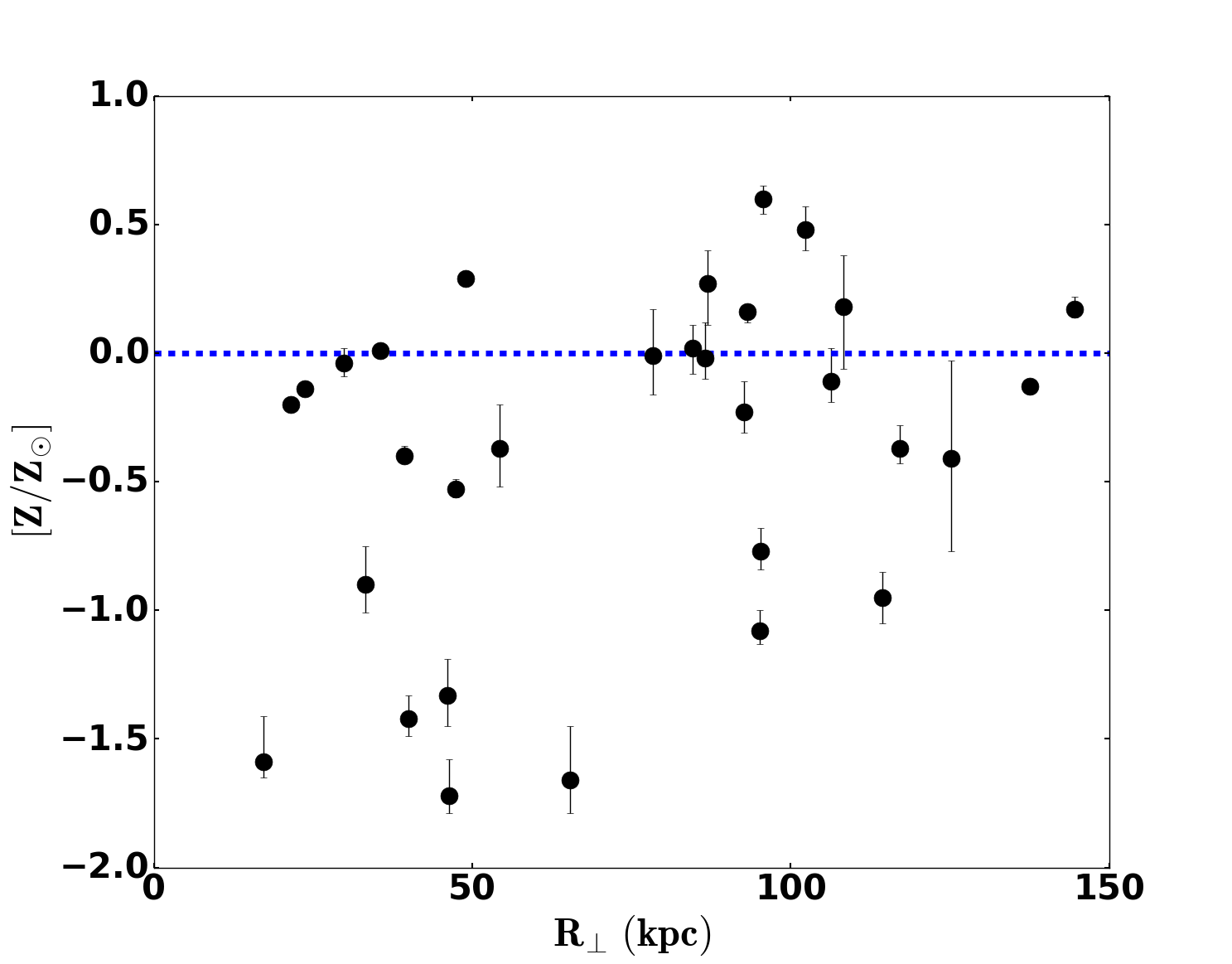}
    \caption{Scatter plot of the median inferred metallicity as a function of the impact parameter. Error bars represents the probability contained between the 34th (lower limit) and the 68th (upper limit) percentile.}
    \label{fig:metvrho}
\end{figure}

\begin{figure}[t]
    \includegraphics[width=1.08\linewidth]{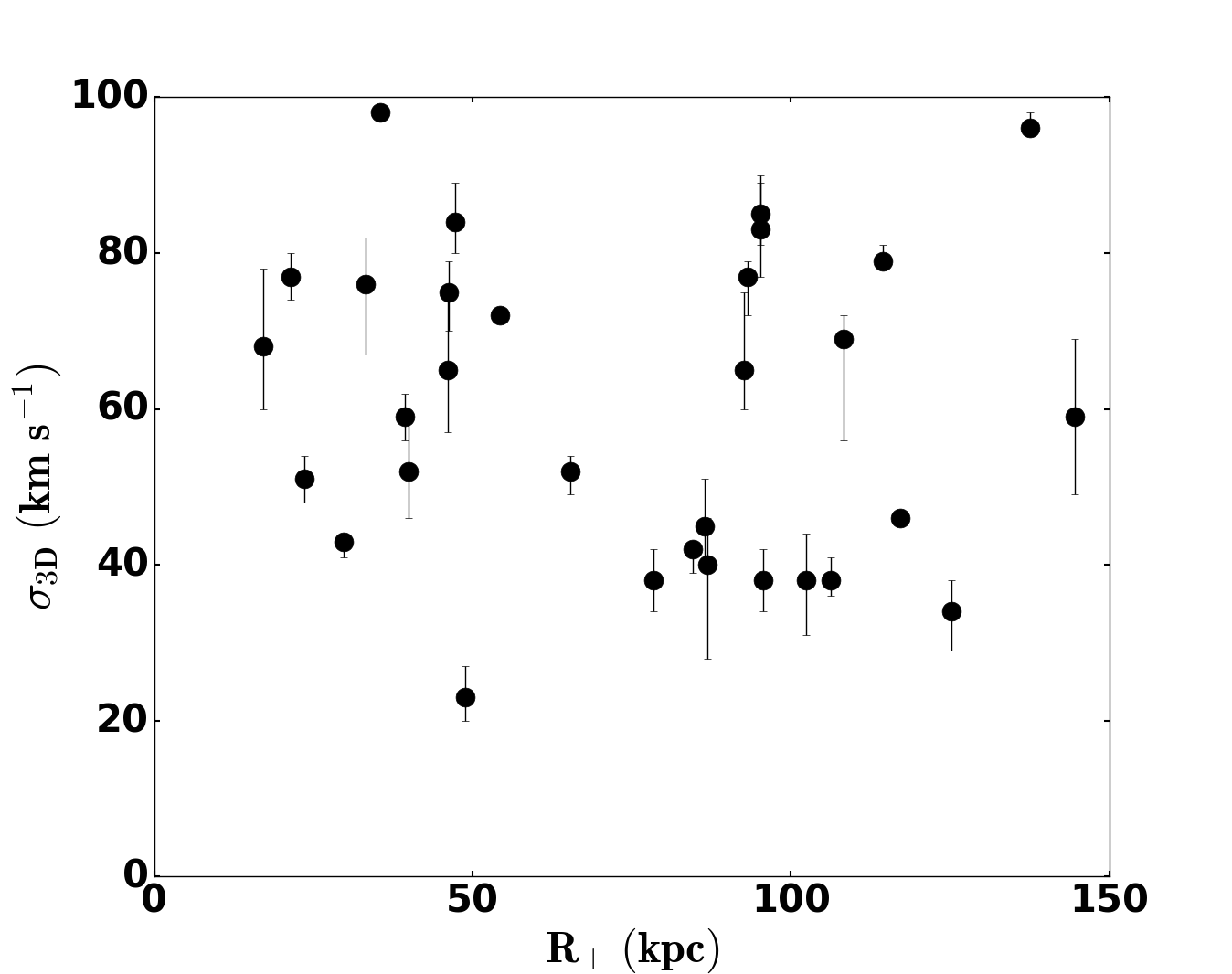}
    \caption{Scatter plot of the median inferred  $\sigma_{\rm 3D}$ as a function of the impact parameter. Error bars represents the probability contained between the 34th (lower limit) and the 68th (upper limit) percentile.}
    \label{fig:turbvrho}
\end{figure}

Beyond the properties of the turbulence, 
similarly to P17, we also find a negative correlation with $>99.99\%$ statistical significance between the metallicity and $N_{\rm H\ I}$. As shown in Figure \ref{fig:metvturb}, 
we also find that the inferred metallicity and the inferred $\sigma_{\rm 3D}$ are negatively correlated with $>97\%$ statistical significance. 

Furthermore, we show the inferred metallicity and $\sigma_{\rm 3D}$ as a function of the impact parameter ($R_{\perp}$) in Figures \ref{fig:metvrho} and \ref{fig:turbvrho} respectively. We find that the metallicity and impact parameter are positively correlated at $>97.9\%$ confidence as compared to the result found in P17, where no correlation was reported. However, $\sigma_{\rm 3D}$ does not appear to show any significant correlation with $R_{\bot}$ as, while high turbulence is found preferentially in the larger \ion{H}{1} columns, both high and low velocities are present throughout the halos of these galaxies.

\section{Discussion and Summary}

Turbulence is  present within the CGM, and promotes the formation of density and temperature gradients that give rise to a multiphase medium. 
We test the ability of models of ionized turbulent media to describe observations of the CGM by conducting an MCMC analysis on the recent COS-Halos data using the MAIHEM non-equilibrium chemistry code that includes turbulence. We also compare these results with those of a previous MCMC study that used PIE models (P17). 

 Generally, we find that most of the MAIHEM fits agree well with observations, and that non-equilibrium turbulent media provide a good fit to both high and low ionization state ions. 
 Indeed, while the low-ions are modeled with comparable success to the PIE case, only with 
 the turbulent MAIHEM models we are able to reproduce simultaneously observations of ions of different ionization potential, including \ion{O}{6}. However, we note that while lower state ions are almost always reproduced, for \ion{O}{6} we are able to find accurate fits  (within 1$\sigma$ of the observations) for about 50\% of the systems that show \ion{O}{6} absorption. 
 
 As noted, our model is not expected to provide a good description of absorbers in which different components arise from different gas phases projected along the line of sight. We therefore investigate in more detail the kinematics of the 13 systems for which \ion{O}{6} is well matched by the MAIHEM model. We find that 11 of these 13 systems in fact  contain broad \ion{O}{6} absorbers that are generally aligned in velocity with low ions, although they appear as broad absorbers with $b >$ 40~km~s$^{-1}$ and linewidths > 30~km~s$^{-1}$. Furthermore, only 4 of these systems show double component absorption of \ion{O}{6} with 2 of those components not matching with low ion absorption within 50~km~s$^{-1}$.
 
Conversely, out of the 15 systems that do not match $N_{\rm OVI}$, 6 show $\approx 2-3$ components in \ion{O}{6}, with a mixture of broad, narrow components ($b<$35~km~s$^{-1}$ and line widths $< 15$~km~s$^{-1}$) matched to low ions, as well as components that are not matched to low ions. This difference suggests a mixed population with absorbers that are potentially multiphase but well-mixed, and absorbers that are  possibly superimposed along the line of sight. Thus, only a subset of these systems may require turbulence to fully capture the multiphase nature of the CGM, while the remaining set the MAIHEM model may still rely on approximations that in fact do not capture the full properties of the halo gas.
 
 When comparing the inferred distribution of metallicity for the entire sample using the MAIHEM and PIE models, we find general agreement although the MAIHEM solutions are more clustered towards higher metallicity, approaching solar values. Inferred properties for absorption line systems that trace the CGM, such as metallicity as studied here, appear therefore subject to non-negligible systematic uncertainties related to the underlying model assumption. 

Furthermore, we find that all of the 32 systems analyzed in this work are fit with solutions that require some amount of turbulence. We find the turbulence of these systems has a possible bimodal distribution with the two maxima at 40 and 75~km~s$^{-1}$. When restricting the analysis to
systems for which MAIHEM provides an excellent fit, we see a clear peak at around 40~km~s$^{-1}$, with a tail extending above 60~km~s$^{-1}$.

Finally, we uncover a positive correlation between $N_{\rm H\ I}$ and $\sigma_{\rm 3D}$ with a $>99.9\%$ confidence, accompanied by a positive correlation between $N_{\rm Si\ II}$ and $\sigma_{\rm 3D}$ at the $>99.9\%$ confidence level. We also see higher metallicity correlated with lower hydrogen column densities, and we find tentative evidence for higher metallicity in lower turbulent gas. 
Furthermore, metallicity is positively correlated at $>97.9\%$ with the impact parameter, while no strong correlation emerges between the turbulent velocity and impact parameter.

As many studies have found $N_{\rm H\ I}$ to decrease with impact parameter \citep[figure 4 of P17 for this sample;][]{tumlinson2013,lehner2013bimodal,savage2014}, the positive correlation between $N_{\rm H\ I}$ and $\sigma_{\rm 3D}$ may be interpreted at first as a suggestion of a more turbulent inner halo, with the degree of turbulence decreasing with impact parameter or column density. An increased turbulence near the center of the halo can be fueled by energy injection from the disk, for example in the form of supernovae or active galactic nuclei   \citep[e.g.][]{2013lilly,2015voit,2015crighton,fox2017,2017muratov,2017reviewARA&A..55..389T}.

However, when examining directly the correlation between turbulent velocity and projected impact parameter, we find that, while sightlines intersecting ($R_\perp \lesssim$ 50~kpc) preferentially show high turbulent velocities, there are also a few systems at high turbulence at large impact parameters. Altogether, therefore, there is no unique evidence of a clear gradient of the turbulent velocity with radius, implying the presence of additional mechanisms that inject energy throughout the entire CGM, such as galactic fountains \citep[e.g.][]{shapiro1976consequences,kahn1994galactic,fraternali2014galactic,voit2018role}.

In a dynamic state where gas is continuously stirred inside the halo, many small cold gas clumps moving at high velocity dispersion would lead to multiple absorption components seen in absorption. When studying a correlation between $\sigma_{\rm 3D}$ and the  number of \ion{Si}{2} absorbers, however, no trend towards more components at higher turbulent velocities is found.
It is therefore more likely that we are observing larger systems, that are turbulent due to direct energy injection from feedback processes or from the condensation of hot gas out of the halo \citep[e.g.][]{peek2008ongoing,fraternali2010gas,joung2012gas}.

Not being a cosmological model, MAIHEM is unlikely to capture 
the full complexity of the CGM, and in particular it cannot capture multiple gas phases that are projected along the line of sight \citep{liang2018observing}, or radial gradients due to different mechanisms operating at different distances from the central galaxies. Nevertheless, our analysis represents a first step in the direction of including more realistic gas hydrodynamic models that extend the commonly used PIE models. Future work that builds on our implementation and extends its result to a cosmological context will be essential to better interpret current observations and fully understand the interactions between the CGM and the central galaxies.

\vspace{0.5in}

We thank Xavier Prochaska for sharing the COS-Halos survey data with us as well as Sanchayeeta Borthakur and Jessica Werk for their advice on turbulence in the inner halo and helping to interpret the resulting correlations. EBII was supported by the National Science Foundation Graduate Research Fellowship Program under Grant No. 026257-001.  MF acknowledges support by the Science and Technology Facilities Council [grant number  ST/P000541/1]. This project has received funding from the European Research Council (ERC) under the European Union's Horizon 2020 research and innovation programme (grant agreement No 757535). ES gratefully acknowledges the Simons Foundation for funding the workshop Galactic Winds: Beyond Phenomenology which helped to inspire this work. He was supported by the NSF under grant AST14-07835 and NASA theory grant NNX15AK82G. This work used the DiRAC Data Centric system at Durham University, operated by the Institute for Computational Cosmology on behalf of the STFC DiRAC HPC Facility (www.dirac.ac.uk\footnote{www.dirac.ac.uk}). This equipment was funded by BIS National E-infrastructure capital grant ST/K00042X/1, STFC capital grants ST/H008519/1 and ST/K00087X/1, STFC DiRAC Operations grant ST/K003267/1 and Durham University. DiRAC is part of the National E-Infrastructure. Simulations presented in this work were carried out  on the NASA Pleiades supercomputer maintained by the Science Mission Directorate High-End Computing program and on the Stampede2 supercomputer at Texas Advanced Computing Center (TACC) through Extreme Science and Engineering Discovery Environment (XSEDE) resources under grant TGAST130021.

\software{FLASH \citep[v4.4]{fryxell2000flash}, CLOUDY \citep{ferland2013}, yt \citep{2011turk}}

\bibliographystyle{yahapj}
\bibliography{main}

\begin{longrotatetable}
\begin{deluxetable}{lccccccccc}
\tablecaption{Results of the MCMC analysis using MAIHEM models. We do not include the following systems from the P17 sample in this analysis because they lack sufficient data to constrain the models: J0226+0015\_268\_22, J0935+0204\_15\_28, J0943+0531\_216\_61, J1133+0327\_164\_21, J1157-0022\_230\_7, J1342-0053\_77\_10, J1437+5045\_317\_38, J1445+3428\_232\_33, J1550+4001\_97\_33, J1617+0638\_253\_39, and J2257+1340\_270\_40. The log $N_{\rm H\ I}$, [Z/H], log U, $\sigma_{\rm 3D}$, log $N_{\rm tot}$ columns list the 68\% c.l interval and the median of the MCMC PDFs. $^{a}$ Flag showing the treatment for $N_{\rm H\ I}$ : $0 =$ Gaussian; $-3 =$ Uniform. $^{b}$ Number of ion detections for model constraints.}
\label{tab:tableall}
\tablehead{
\colhead{System}& \colhead{Redshift} & \colhead{log $N_{\rm H\ I\ meas.}$} & \colhead{f$^{a}$} & \colhead{Ion$^{b}$} & \colhead{log $N_{\rm H\ I}$} & \colhead{[Z/H]} & \colhead{log $U$} & \colhead{$\sigma_{\rm 3D}$} & \colhead{log $N_{\rm tot}$}\\
&  & \colhead{(cm$^{-2}$)} &  &  & \colhead{(cm$^{-2}$)} & & & \colhead{(km~s$^{-1}$)} & \colhead{(cm$^{-2}$)}}
\startdata
J0401-0540\_67\_24  & 0.22  & $15.45 \pm 0.03$ & 0    & 11 & 15.44,15.45,15.45 & -0.10,-0.02,0.12 & -2.00,-1.90,-1.83 & 40,45,51 & 18.08,18.24,18.64  \\
J0803+4332\_306\_20 & 0.25  & $14.78 \pm 0.04$ & 0    & 10 & 14.77,14.78,14.80 & -0.16,-0.01,0.17 & -1.62,-1.45,-1.27 & 34,38,42 & 17.77,17.89,18.04  \\
J0910+1014\_34\_46  & 0.14  & $17.25 \pm 1.25$ & -3   & 9 & 18.08,18.20,18.34 & -1.05,-0.95,-0.85 & -1.83,-1.76,-1.70 & 78,79,81 & 19.58,19.67,19.74  \\
J0910+1014\_242\_34 & 0.26  & $16.58 \pm 0.06$ & 0    & 9 & 16.61,16.63,16.66 & 0.14,0.17,0.22 & -2.22,-2.19,-2.16 & 49,59,69 & 18.93,18.95,18.99    \\
J0914+2823\_41\_27  & 0.24  & $15.45 \pm 0.03$ & 0 & 9 & 15.44,15.45,15.46 & -0.19,-0.11,0.02 & -1.82,-1.78,-1.76 & 36,38,41 & 18.07,18.18,18.27   \\
J0925+4004\_196\_22 & 0.25  & $19.55 \pm 0.15$ & 0    & 10 & 19.65,19.71,19.78 & -0.08,0.02,0.11 & -2.67,-2.58,-2.50 & 39,42,43 & 18.01,18.08,18.22 \\
J0928+6025\_110\_35 & 0.15  & $19.35 \pm 0.15$ & 0    & 10 & 19.27,19.33,19.41 & -0.84,-0.77,-0.68 & -3.73,-3.56,-3.45 & 77,83,89 & 19.29,19.71,20.24 \\
J0943+0531\_106\_34 & 0.23  & $16.03 \pm 0.54$ & 0    & 10 & 16.08,16.31,16.56 & -0.77,-0.41,-0.03 & -1.92,-1.65,-1.42 & 29,34,38 & 17.96,18.23,18.69  \\
J0950+4831\_177\_27 & 0.21  & $18.20 \pm 0.30$ & -3   & 11 & 18.30,18.36,18.42 & -1.13,-1.08,-1.00 & -2.64,-2.60,-2.56 & 81,85,90 & 19.37,19.48,19.63   \\
J1009+0713\_170\_9  & 0.36  & $18.50 \pm 0.50$ & -3   & 9 & 18.03,18.06,18.11 & -0.56,-0.53,-0.49 & -2.29,-2.28,-2.26 & 80,84,89 & 20.22,20.28,20.38   \\
J1009+0713\_204\_17 & 0.23  & $17.50 \pm 1.00$ & -3   & 9 & 17.96,18.11,18.25 & -1.79,-1.66,-1.45 & -1.95,-1.94,-1.92 & 49,52,54 & 20.27,20.33,20.44   \\
J1016+4706\_274\_6  & 0.25  & $17.10 \pm 0.02$ & 0    & 11 & 17.09,17.09,17.10 & -0.16,-0.14,-0.12 & -2.12,-2.10,-2.08 & 48,51,54 & 19.31,19.37,19.45  \\
J1016+4706\_359\_16 & 0.17  & $17.50 \pm 1.00$ & -3   & 8 & 18.28,18.37,18.43 & -1.79,-1.72,-1.58 & -2.06,-2.03,-2.00 & 70,75,79 & 20.26,20.29,20.32   \\
J1112+3539\_236\_14 & 0.25  & $16.70 \pm 0.90$ & -3   & 9 & 16.97,17.13,17.28 & -0.52,-0.37,-0.20 & -2.63,-2.56,-2.47 & 71,72,73 & 18.74,18.79,18.84   \\
J1133+0327\_110\_5  & 0.24  & $18.60 \pm 0.06$ & 0    & 9 & 18.57,18.60,18.63 & -1.65,-1.59,-1.41 & -2.46,-2.40,-2.33 & 60,68,78 & 19.77,20.16,20.29  \\
J1220+3853\_225\_38 & 0.27  & $15.88 \pm 0.06$ & 0    & 8 & 15.85,15.88,15.91 & 0.29,0.49,0.68 & -2.12,-2.08,-2.05 & 28,33,38 & 17.69,17.83,17.96   \\
J1233-0031\_168\_7  & 0.32  & $15.57 \pm 0.02$ & 0    & 10 & 15.55,15.56,15.58 & -0.09,-0.04,0.02 & -1.84,-1.81,-1.78 & 41,43,44 & 18.13,18.23,18.33  \\
J1233+4758\_94\_38  & 0.22  & $16.74 \pm 0.04$ & 0    & 10 & 16.72,16.74,16.76 & -0.15,-0.13,-0.11 & -2.04,-2.04,-2.03 & 95,96,98 & 19.38,19.42,19.49  \\
J1241+5721\_199\_6  & 0.21  & $18.15 \pm 0.35$ & -3   & 12 & 17.83,17.84,17.86 & -0.22,-0.20,-0.18 & -2.36,-2.36,-2.35 & 74,77,80 & 19.97,20.00,20.04  \\
J1241+5721\_208\_27 & 0.22  & $15.29 \pm 0.07$ & 0    & 9 & 15.28,15.30,15.32 & 0.54,0.60,0.65 & -1.70,-1.68,-1.66 & 34,38,42 & 18.42,18.53,18.77   \\
J1245+3356\_236\_36 & 0.19  & $14.76 \pm 0.04$ & 0    & 10 & 14.75,14.76,14.78 & -0.43,-0.37,-0.28 & -2.01,-1.94,-1.81 & 46,46,47 & 17.97,18.07,18.18   \\
J1322+4645\_349\_11 & 0.21  & $17.14 \pm 0.03$ & 0    & 11 & 17.12,17.13,17.14 & -0.43,-0.40,-0.36 & -2.20,-2.17,-2.14 & 56,59,62 & 19.57,19.67,19.77  \\
J1330+2813\_289\_28 & 0.19  & $17.03 \pm 0.12$ & 0    & 12 & 16.25,16.27,16.34 & 0.12,0.16,0.19 & -1.91,-1.89,-1.87 & 72,77,79 & 19.80,19.83,19.87  \\
J1342-0053\_157\_10 & 0.23  & $18.50 \pm 0.50$ & -3   & 11 & 18.54,18.55,18.57 & 0.00,0.01,0.02 & -2.78,-2.77,-2.76 & 97,98,99 & 20.72,20.75,20.78  \\
J1419+4207\_132\_30 & 0.18  & $16.63 \pm 0.19$& 0    & 11 & 16.73,16.85,16.92 & -0.31,-0.23,-0.11 & -2.16,-2.13,-2.11 & 60,65,75 & 19.23,19.27,19.37  \\
J1435+3604\_68\_12  & 0.20  & $19.80 \pm 0.10$ & 0    & 9 & 19.75,19.79,19.85 & -1.49,-1.42,-1.33 & -2.76,-2.74,-2.72 & 46,52,59 & 19.71,19.95,20.19   \\
J1435+3604\_126\_21 & 0.26  & $15.25 \pm 0.06$ & 0    & 10 & 15.24,15.27,15.29 & 0.11,0.27,0.40 & -1.95,-1.83,-1.79 & 28,40,46 & 17.89,18.11,18.38  \\
J1514+3619\_287\_14 & 0.21  & $17.50 \pm 1.00$ & -3   & 8 & 18.10,18.21,18.32 & -1.45,-1.33,-1.19 & -3.79,-3.71,-3.62 & 57,65,74 & 18.93,19.02,19.15   \\
J1550+4001\_197\_23 & 0.31  & $16.50 \pm 0.02$ & 0    & 10 & 16.50,16.50,16.51 & -0.06,0.18,0.38 & -2.69,-2.63,-2.56 & 56,69,72 & 18.77,19.06,19.21  \\
J1555+3628\_88\_11  & 0.19  & $17.17 \pm 0.20$ & 0    & 10 & 17.50,17.60,17.69 & -1.01,-0.90,-0.75 & -2.15,-2.12,-2.09 & 67,76,82 & 20.00,20.23,20.27  \\
J1619+3342\_113\_40 & 0.14  & $14.96 \pm 0.03$ & 0    & 9  & 14.95,14.95,14.97 & 0.40,0.48,0.57 & -2.19,-2.02,-1.90 & 31,38,44 & 17.47,17.61,17.96  \\ 
J2345-0059\_356\_12 & 0.25  & $16.00 \pm 0.04$ & 0    & 9 & 15.98,15.99,16.01 & 0.27,0.29,0.32 & -2.68,-2.63,-2.60 & 20,23,27 & 17.57,17.68,17.79 \\
\enddata
\end{deluxetable}
\end{longrotatetable}

\end{document}